\documentclass{aa}  
\usepackage{graphicx}
\usepackage{epsfig}
\usepackage{txfonts}
\begin{document}
   \title{Bar-driven injection of intergalactic matter into galactic
   halos}

   \subtitle{}

   \author{M. L\'opez-Corredoira\inst{1}}

   \offprints{martinlc@iac.es}

\institute{
$^1$ Instituto de Astrof\'\i sica de Canarias, C/.V\'\i a L\'actea, s/n,
E-38200 La Laguna (S/C de Tenerife), Spain}

   \date{Received xxxx; accepted xxxx}

 
  \abstract
   {}
   {The non-conservative gravitational potential of barred galaxies,
   or of any other non-axisymmetric structure, produces a loss of
   energy in  infalling particles of the intergalactic medium
   into the galaxy,  which are trapped in its potential. 
   This dynamical friction can contribute towards increasing the total mass 
   of  barred galaxies.}
   {Analytical calculations of the energy loss are carried out using
   the orbits of the particles derived numerically.
   Theoretical predictions are compared with  observations
   through the statistical analysis of the rotation curves of barred 
   and non-barred galaxies, either in cluster or field galaxies.}
   {There is a net effect of accretion, but it is normally very low
   in relative terms. It is only significant ($>10$\% of the total
   mass of the dark matter halo in the life of the galaxy) if the
   density of the intergalactic medium is higher than $\sim 3\times 10^{13}$ 
   M$_\odot $/Mpc$^3$ (or considerably lower in cases 
   of motions of the galaxies 
   close to the IGM average motion, or perpendicular to 
   the plane of the galaxy, or when the halo mass is low).
   
   Data on rotation curves do not show clear trends towards higher
   halo mass for barred galaxies, only slight trends 
   for early-type spiral galaxies.
   In any case, the statistical uncertainties are limited to the detection of
   differences in masses $>\approx $20\%,
   so the effect of bar-driven injection of intergalactic matter into 
   galactic halos might be present with relative contributions
   to the average mass of these barred galaxies lower than 20\%.}
   {}

\titlerunning{Bar-intergal.}
\authorrunning{L\'opez-Corredoira}

   \maketitle
%

\section{Introduction}

Intergalactic matter (IGM) is still a field of astrophysics with many open 
questions and few answers. There are either theoretical arguments
or interpretations of observations that suggest the existence of 
an IGM, although the definitive details of important
parameters such as the mean density, whether  continuous or in
the form of clouds, etc. are not clear yet.
Cold dark matter models predict substructure within 
galactic- and cluster-mass halos 
that form in a hierarchical accretion Universe (Klypin et al. 1999; 
Moore et al. 1999) that could constitute some of the clumpy structures in
the IGM. Indeed, 
stellar and cold gas in galaxies contribute  $8 (^{+4}_{-5})$\%  
of the total amount of the Big Bang-produced baryonic matter predictions 
(Bell et al. 2003); this implies a low overall efficiency of galaxy formation,
and that the rest of the baryonic material must be some way away from the
visible galaxies, possibly in the IGM.
Some interstellar gas is ejected into the IGM by 
tidal interactions between galaxies (Morris \& van den Bergh 1994), 
SN explosions, etc. Of course, the existence
of intracluster gas in rich and irregular clusters, 
which have diffuse emission and X-ray emission associated with 
hot gas, is well established. Possibly, some high-velocity clouds (HVCs) 
observed in the radio are located outside our Galaxy, i.e. 
belonging to the IGM (Blitz et al. 1999; Braun \& Burton 1999). 
There is also a large number of lines visible in optical at high redshifts 
or UV in low redshifts---the Lyman-$\alpha $ forest---which are presumably 
caused by clouds along the line of sight of QSOs (Rauch 1998).

There is a need for a substantial accretion of low 
angular momentum material from the IGM into the galaxies (Fraternali et al. 
2007). Chemical evolution in the local Galactic disc (the G-dwarf problem,
$^9$Be abundance and others; Rocha-Pinto \& Maciel 1996; Casuso \& Beckman 
1997) needs a continual but episodic infall of 
metal-poor gas that mixes slowly with the rest of 
the interstellar medium. L\'opez-Corredoira et al. (1999) deduce that 
the IGM mass must currently
represent at least around one half of the total mass of the Local Group, 
given that the accretion rate, as inferred from chemical evolution, has not 
decreased significantly during the disc lifetime.

The existence of matter in the IGM should have some effects
on the formation and evolution of the galaxies. For instance,
it is a possible explanation for the formation of warps in spiral
galaxies (L\'opez-Corredoira et al. 2002a). It is also, as said,
a likely explanation for the chemical evolution of  galaxies
(Casuso \& Beckman 1997). Here, we study another effect:
the accretion of this IGM matter due to dynamical friction.
It is clear that the disc mass might be fed  by the 
collisional friction of baryonic matter with  disc gas.
The enrichment of the halo by means of a gravitational interaction
is instead the topic to be treated here. In particular, the
component of the galaxy responsible of this gravitational interaction
will be the bar.

Bar--halo interactions have been studied for several reasons.
For instance, as a mechanism of angular momentum transfer.
It is well established that strong bars rotating in dense halos generally
slow down as they lose angular momentum to the halo through dynamical 
friction (Debattista \& Sellwood 1998, 2000; Athanassoula 2003, 2005). 
This friction can be avoided or found to be anomalously weak in some 
circumstances and bar slowdown 
can be delayed for a period in a metastable state
(Sellwood \& Debattista 2006), although Sellwood \& Debattista
(2006) demonstrate that mild external, or internal, perturbations quickly 
restore the usual frictional drag; it is therefore unlikely  that a
strong bar in a galaxy having a dense halo could rotate for a long period 
without friction. A more
long-lasting effect to be considered is that the velocity dispersion of the
halo particles is high enough to stop the resonances from absorbing
considerable amounts of angular momentum (Athanassoula 2003), hence
prohibiting bar slowdown. Owing to this angular momentum transfer and 
self-consistent re-equilibration, strong realistic bars will modify the 
cusp profile, lowering the central densities within about 30\% 
of the bar radius in a few bar orbits (Weinberg \& Katz 2006). 

In this paper, another effect of the bar will be explored, 
also using the dynamical friction of this kind of gravitational interaction: 
the influence of the bar to produce the accretion of IGM
matter (initially non-gravitationally linked to the galaxy)
into the halo. 
That is, the interaction with the bar will produce
the loss of energy of some particles that cross the galaxy
and that will be trapped in it, since they do not have escape velocity, to form part of the galactic halo.
This mechanism applies to both baryonic and non-baryonic
matter because the friction is purely gravitational.
In \S \ref{.varene}   this loss
of energy and the amount of accreted matter depending
on  IGM parameters for a Milky-Way-like galaxy is calculated semianalytically.
In \S \ref{.other}, it is explained that other factors of the non-conservative
potential, for example  approaches to individual stars, produce
a much lower effect that is totally negligible.
In \S \ref{.observations}, statistics with observational data
are performed in order to see whether barred galaxies have
larger masses.

\section{Variations of energy in a barred potential}
\label{.varene}

We analyse the variations of the energy in particles
that follow orbits in certain kinds of non-conservative potentials:
those in which the gravitational force varies with time.
This happens, for instance, in a barred galaxy, because the
rotation of a non-axisymmetric structure produces variations
in the potential due to the bar with respect to an inertial
frame. However, the rotating disc has
a stationary potential since the axial distribution of mass
does not change. If the bar rotates with angular velocity $\omega$,
the variation in the energy of a particle along
its path (orbit) is:

\begin{equation}
\Delta E=\int _{orbit} \vec{F}(R,\phi-\omega t,z)d\vec{r}
\end{equation}\[=
\int _{orbit}\vec{F}(R,\phi-\omega t,z)(dR\vec{e_R}+Rd\phi \vec{e_\phi }
+dz\vec{k})
,\]
in cylindrical coordinates, where $\vec{F}$ is the gravitational
force produced by the bar. 
If we change to the variable $\phi '=\phi -\omega t$ (that is,
we adopt the frame that is moving with the bar):

\begin{equation}
\Delta E=\int _{orbit}\vec{F}(R,\phi',z)
(dR\vec{e_R}+Rd\phi '\vec{e_{\phi '}}+dz\vec{k}+R\omega dt \vec{e_{\phi '}})
.\end{equation}

If the orbits of particles begin at infinite distance and finish at
infinite distance:

\begin{equation}
\int _{r=\infty,\ orbit}^{r=\infty}\vec{F}(\vec{r'})d\vec{r'}=0
{\rm \ (because\ it\ does\ not\ depend\ on\ time)}
,\end{equation}
\begin{equation}
\Delta E=\omega \int _{r=\infty, orbit}^{r=\infty}F_{\phi '}(R,\phi ',z)
Rdt =\omega \Delta J_z
\label{deltaE}
.\end{equation}

The Jacobi integral 
\begin{equation}
H=E-\omega J_z
\end{equation}
is conserved.

Let us consider a monodimensional bar. This is an approximate description
of  real 3D-bars because the major axis is considerably larger than the other
axis---otherwise, they would not be bars---as can be observed, for instance, 
in the galaxies NGC 1300, NGC 7479, NGC 4123, NGC 1433, NGC 4999, etc. 
(Sandage \& Bedke 1994) or in our own Galaxy (L\'opez-Corredoira
et al. 2001, 2007). The mass density distribution will be $\lambda (l)$ as a function of the 
distance $l$ to its centre. The azimuthal component of the gravitational
force produced by this bar of radius $R_0$ on a particle of mass $m$ will be:

\begin{equation}
F_\phi (R,\phi ',z)=-Gm\int _{-R_0}^{R_0}dl\frac{\lambda (l)l\sin \phi '}
{(z^2+R^2+l^2-2Rl\cos \phi ')^{3/2}}  
\label{Fphi}
\end{equation}

If we assume $\lambda (l)={\rm constant}=\lambda $, the integral can be solved:

\begin{equation}
F_\phi [\vec{r}'(R,\phi ',z)]=\frac{-Gm\lambda \sin \phi '}{z^2+R^2(1-\cos ^2\phi ')}
\end{equation}\[
\times \left(r_+-\frac{R_0^2}{r_+}+\frac{R_0R\cos \phi '}{r_+}-r_-+\frac{R_0^2}
{r_-}+\frac{R_0R\cos \phi '}{r_-}\right)
,\]
\[r_+\equiv |\vec{r}'-\vec{R_0}|;
r_-\equiv |\vec{r}'+\vec{R_0}|; \vec{R_0}\equiv R_0\vec{i}
.\]

Therefore, if we want to calculate the gain/loss of energy of the particle,
we must just know its orbit $\vec{r}(t)$ and use expressions (\ref{deltaE})
and (\ref{Fphi}) to calculate $\Delta E$.

\subsection{Numerical calculation of orbits}

For our problem, we  use 
semianalytical calculations. The solution of eqs. (\ref{deltaE}),
(\ref{Fphi}) is calculated numerically, and the orbit of each
particle itself is not solved numerically as an outcome of
a numerical simulation in which all particles interact with each other
but as the numerical integration of its motion equation. 
Indeed, we can consider the gravitational potential
of the infalling particles as due only to the components of the galaxies
and neglect the interaction due to other IGM particles.

Analytically, we cannot calculate the orbit of a particle in a
galactic potential; the integrals are too complex to find the solution.
But we can calculate $\vec{r}(t)$ simply by numerical calculation
of the differential equation $\ddot{\vec{r}}=\vec{a}_{galaxy}$, the
total acceleration produced by the total potential of the galaxy.
It is then straightforward to obtain $\vec{r}'(t)$ (the position in
the frame that is moving with the bar) from $\vec{r}(t)$
by using $\phi '=\phi-\omega t$.  
In this case, of course, for the calculation of the orbit, all the potentials
are to be taken into account not only the potential of
the bar. We can approximate this potential as the sum of several
components:

\begin{equation}
\vec{a}_{galaxy}=\vec{a}_{\rm halo}+\vec{a}_{\rm disc}+\vec{a}_{\rm bar}
.\end{equation}
We  give analytical expressions for these acceleration  
terms as integrals. All these integrals will be  performed 
numerically with the Simpson algorithm, and the solution to 
the differential equation $\ddot{\vec{r}}=\vec{a}_{galaxy}$
will be carried out with a variable step depending
on the acceleration:
\begin{equation}
\Delta t=\frac{50 s}{a({\rm m/s^2})+10^{-12}}
.\end{equation}

The Jacobi integral $H$ is conserved for each orbit
within variations of $\sim 0.1$\% .
The accuracy in the orbit calculation was also checked by testing
that the orbit is approximately the same when decreasing this step 
$\Delta t$ by a factor of 50 ($\Delta t=\frac{1 s}{a({\rm m/s^2})+10^{-12}}$): 
the difference in the position of both calculations after a travelled
path of more than 1000 kpc (once the particle fell  from a distance $r_0$
towards the centre and, after that, it 
crosses  the shell at $r=r_0$; with
$r_0=400-500$ kpc) is $\sim 0.4$ kpc on average. 
These errors  are not important for the present 
calculations because we do not require too much accuracy in the orbit;
it is only used for the path integral of eq. (\ref{deltaE}).
Moreover, a small random error added to the orbit is equivalent to a 
small random variation of the initial conditions, which in comparison
with the range of impact parameters (0-200 kpc) is negligible.

\subsubsection{Halo potential}

We approximate the halo as spherical, i.e. we neglect the ellipticity:

\begin{equation}
\vec{a}_{\rm halo}(\vec{r})=-\frac{[4\pi G\int _0^rdt\ t^2
\rho _{halo}(t) ]\vec{r}}{r^3}
.\end{equation}

We then use a truncated flat model by Battaglia et al. (2005):

\begin{equation}
\rho _{halo}(r)=\frac{M_{\rm halo}}{4\pi}\frac{a^2}{r^2(r^2+a^2)^{3/2}}
\label{halodens}
.\end{equation} 

\subsubsection{Disc potential}

For the disc, we use a typical exponential function considering
 its thickness to be negligible. In order to reduce the
calculation to only one integral, which reduces  the calculation time very considerably,
we express it in elliptic integrals [Binney \& Tremaine 1987, eq.
(2-142b)]:

\begin{equation}
\vec{a}_{\rm disc}[\vec{r}(R,\phi ,z)]=-\frac{\partial \phi _{\rm disc}}
{\partial R}\vec{e_R}- \frac{\partial \phi _{\rm disc}}{\partial z}\vec{k}
,\end{equation}
\begin{equation}
\phi _{\rm disc}=-\int _0^\infty dT\ T \int _0^{2\pi }d\alpha \frac{G\sigma (T)}
{\sqrt{z^2+R^2+T^2-2RT\cos (\phi -\alpha)}}
\end{equation}\[=
\frac{-2G}{\sqrt{R}}\int _0^\infty dT\ K(k)k\sigma (T)\sqrt{T}
,\] 
where
\begin{equation}
k(R,T,z)=\sqrt{\frac{4RT}{z^2+(R+T)^2}}
,\end{equation}
and $K(k)$ are the complete elliptic integrals of the first kind,
\begin{equation}
K(k)=\int _0^1\frac{dt}{\sqrt{(1-t^2)(1-k^2t^2)}}
.\end{equation}
The two derivatives of $\phi $ are:

\begin{equation}
\frac{\partial \phi _{\rm disc}}{\partial R}=
\frac{G}{R^{3/2}}\int _0^\infty dT 
\end{equation}\[
\times \left[K(k)-\frac{1}{4}\left(\frac{k^2}
{1-k^2}\right)\left(\frac{T}{R}-\frac{R}{T}+\frac{z^2}{RT}\right)
E(k)\right]k\sigma (T)\sqrt{T}
\]
(Binney\ \& Tremaine\ 1987,\ eq. [2-146]), and
\begin{equation}
\frac{\partial \phi _{\rm disc}}{\partial z}=
\frac{Gz}{2R^{3/2}}\int _0^\infty dT \frac{k^3E(k)}{1-k^2}
\frac{\sigma (T)}{\sqrt{T}}
,\end{equation}
where $E(k)$ are the complete elliptic integrals of the second kind,
\begin{equation}
E(k)=\int _0^1dt\sqrt{\frac{1-k^2t^2}{1-t^2}}
.\end{equation}

We assume an exponential disc such that: 

\begin{equation}
\sigma (R)=\sigma _0\exp{\left(-\frac{R}{h_R}\right)}
.\end{equation}

\subsubsection{Bar potential}

\begin{figure}
\begin{center}
\vspace{1cm}
\mbox{\epsfig{file=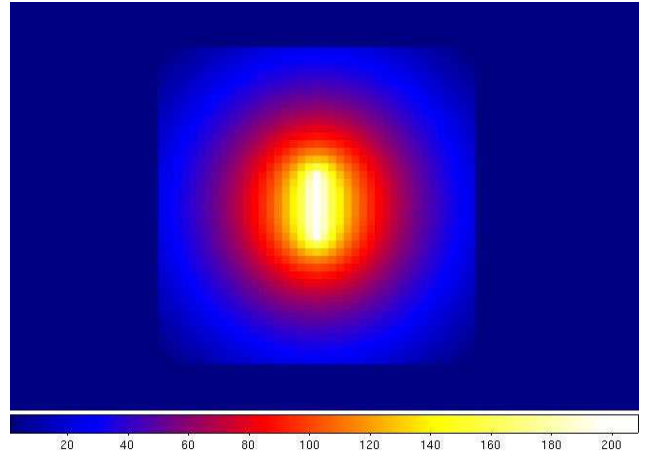,height=6cm}}
\end{center}
\caption{Modulus of the acceleration in a monodimensional bar potential
with the parameters of \S \ref{.parameters}.
Logarithmic scale. Range of x and y between $-$20 kpc and 20 kpc.}
\label{Fig:ds9}
\end{figure}

For the bar, as said, we adopt a monodimensional structure
rotating with angular velocity $\omega $:

\begin{equation}
\vec{a}_{\rm bar}=-G\int _{-R_0}^{R_0}dl\frac{\lambda (l)[\vec{r}-l[\cos 
(\omega t) \vec{i}+\sin (\omega t)\vec{j}]]}
{(z^2+R^2+l^2-2Rl\cos (\phi -\omega t))^{3/2}}  
.\end{equation}
Other bar potentials could be used (e.g., Athanassoula 1992) but
this one is used for its simplicity. The IGM particles will
never cross this 1D structure, because there is zero probability
that the path by chance crosses the line of the bar, so this
potential will not give problems of divergence.
The potential also behaves well at large distance, as shown
in Fig. \ref{Fig:ds9},  the
acceleration falling as $\propto r^{-2}$. 

\subsubsection{Parameters for our numerical calculations}
\label{.parameters}

The parameters that we used are those more
or less typical for Milky Way-like galaxies. 

A total mass for the halo
$M_{\rm halo}=1.2\times 10^{12}$ M$_\odot $, and $a=105$ kpc
(Battaglia et al. 2005).
The mass within a radius $r$ would be $M(r)=M_{\rm halo}\frac{r}{\sqrt{r^2+
a^2}}$.  96.7\% of the halo mass is within $r<400$ kpc, so it can be
approximated for $r>400$ kpc as a Keplerian potential of mass equal to 
$M_{\rm halo}$ (plus the mass of the disc and the bar). 

For the disc, we adopt $\sigma _0=1$ kg/m$^2$ (in agreement with the
solar neighbourhood measurements by Kuijken \& Gilmore 1989) and
$h_R=3.4$ kpc (a typical value in a Milky Way-like galaxy;
L\'opez-Corredoira et al. 2002b). The corresponding total mass of
the disc is $3.5\times 10^{10}$ M$_\odot $.

For the bar, we assume a total mass of $1.7\times 10^{10}$ M$_\odot$
(Sevenster et al. 1999), more
or less in agreement with Kuijken \& Dubinski's (1995) self-consistent
models of the Galaxy. This gives a constant distribution 
$\lambda (l)=1.4\times 10^{20}$ kg/m for a radius $R_0=4$ kpc 
(estimated value of the long-bar radius in our Galaxy, 
L\'opez-Corredoira et al. 2001, 2007). 
This mass of the long bar is indeed not well known (L\'opez-Corredoira
et al. 2007) and the bar of Sevenster et al. (1999) is 35\% shorter
(the axial ratio difference is not so important, as we will see below)
than L\'opez-Corredoira et al. (2007) bar, so we are not setting
 the parameters of the bar accurately, but as a more or less correct order of magnitude 
(see further discussion in \S \ref{.barstrength} of
what would happen with more or less massive bars).

The angular rotation velocity, $\omega $, is taken 
as $1.9\times 10^{-15}$ s$^{-1}$ (as  measured
in the Milky Way; Debattista et al. 2002). Most strong bars in real
galaxies appear to rotate rapidly (Aguerri et al. 2003), so it is
not expected that this velocity will slow down too much.
Although the dynamical  halo--bar friction predicts a quick 
slowdown in this velocity (Debattista \& Sellwood 1998, 2000; Athanassoula
2003, 2005), there
might be a metastable state in which the bar--halo also remains a long time
without friction  (Sellwood \& Debattista 2006), or a state in
which the velocity dispersion of the
halo particles is high, so as to stop the resonances from absorbing
considerable amounts of angular momentum (Athanassoula 2003), thus
prohibiting bar slow down. Whatever 
 the explanation, the fact of fast bars in most
barred galaxies remains, and we therefore
do not consider any friction or change in $\omega $ with time.

With these parameters, the stationary (Lagrangian) 
points of the Galactic potential are at:
L1,L2: $\phi _L'=0, \pi$, $z=0$,
$R_L=5.25$ kpc; L4,L5: $\phi _L'=\pi /2, 3\pi /2$, $z=0$, $R_L=4.59$ kpc
(for more information about the Lagrangian points in a
barred potential see Pfenninger 1990 or Athanassoula 1992), 
whose Jacobi integrals are:
$H=-1.41\times 10^{11}$ J and $-1.29\times 10^{11}$ J respectively.
The fact that the Lagrangian points L4,L5 are 13\% closer to the
centre than the Lagrangian points L1,L2 is expected for average strong
bars; the difference is typically 6\% for average bar models according
to Athanassoula (1992) but it may be as large as 20\% for the strongest bars.
The strength of the bar can be compared with another parameter defined
by Athanassoula (1992): half of the quadrupole moment along its
major axis 
\begin{equation}
Q_m=\frac {1}{5}M_{\rm bar}R_0^2
.\end{equation}
This expression is derived from eq. (5) of Athanassoula with $a=R_0$,
$b=0$ (infinite axial ratio a/b), $n=0$ (constant density along the bar).
With our parameters we get $Q_m=5.4\times 10^4$ (in units $10^6$M$_\odot $
kpc$^2$, as in Athanassoula 1992). This is again an indication that
the strength of our bar is a typical one within strong bars
(weak bars have $Q_m\sim 10^4$, very strong bars have $Q_m\sim 10^5$;
Athanassoula 1992). It is also remarkable that the infinite axial ratio
a/b in a monodimensional bar does not significantly alter  the effects
of a bar with finite but high axial ratios since $Q_m\propto (a^2-b^2)$
(Athanassoula 1992, eq. 5) and $(a^2-b^2)\approx a^2$ for high $a/b$,
either finite or infinite.     

\subsubsection{Initial conditions of the IGM particles}
\label{.initial}

We assume now that our galaxy is moving with respect to the IGM 
(or the IGM is moving with respect to the galaxy).
This hypothesis was already used in L\'opez-Corredoira et al. (2002a)
as an explanation to produce warps in disc galaxies and now will be
explored just to calculate the fraction of the particles that are
gravitationally trapped in the galactic potential due to a negative
energy variation. Basically, the model consists of a constant
density ($\rho _\infty $) medium which has at infinite distance from the
galaxy an average velocity $\vec{v_\infty}$
with respect to the centre of galaxy, and
an angle of $\vec{v_\infty}$ with respect to the plane of the galaxy 
equal to $\theta _\infty $. We neglect the effect of the
dispersion of velocities with respect to this average velocity.

We randomly distribute  the values of 
$b$ (impact parameter with respect to the centre
of the galaxy), $\phi _b$ (azimuthal angle in cylindrical coordinates
with the vector $\vec{v_\infty}$ as vertical axis) following the
conditions for a homogeneous distribution, equivalent to do integrals
in the way $\int _0^{2\pi } d\phi _b\int _0^{b_{max}} db\ b\ (...)$.
We set $b_{max}=200$ kpc. That
is, we neglect the contribution of particles whose impact parameter is
larger than 200 kpc; this is justified because, as we shall see in following
subsections, the particles which lose more energy have lower value of $b$,
except in cases with low $v_\infty $.

Since an infinite distance cannot be introduced in the computer 
(integration of the orbit would take an infinite time), we must 
begin the integration from a finite distance, 
which we choose to be at 400 kpc$<r_0<500$ kpc. We
distribute our particles randomly at a shell with this range of radii.
The distribution with different $r_0$ is useful to integrate over 
different positions of the bar
when the particle reaches the galactic disc; that is, it is equivalent
to doing an integration with different initial bar angles $\omega (t_0)$.
The time to cross 100 kpc at 
around 70 km/s (the velocity of the particle at radius $r_0$ assuming
$v_\infty=50$ kpm/s) is 1.4 Gyr, so doing the integration 
between 400 kpc$<r_0<500$ kpc
is equivalent to doing an integration over a time of 1.4 Gyr, equivalent
to 14 cycles of the bar (one cycle/100 Myr); moreover, there is an 
added dispersion in the position of the bar when the particle reaches
the centre of the galaxy because the different values of $\phi _b$ and $b$
get different infall times to reach the centre, so in practice
we are integrating over a range larger than 1.4 Gyr.
This ensures that all angles of the bar are covered more or less
uniformly (departures from uniformity are negligible and less
important than Poissonian  or other fluctuations).
In this way, we are doing a Monte Carlo simulation over the
variables $b$, $\phi _b$ and $r_0$.

At the distance of 400--500 kpc the non-monopolar terms of the gravitational 
potential are negligible and almost the entire mass of the galaxy is within 
this radius, so the potential can be approximated as one produced by
a point-like source in the centre of the galaxy with its total mass.
Nonetheless, this potential is not totally negligible at these
distances, so we must correct the initial conditions with respect
to the conditions at infinite distance:

\begin{equation}
\vec{r_0}=-\sqrt{r_0-r_{imp}^2}\vec{e}_{v_\infty}+r_{imp}
\vec{e}_{b_\infty}
,\end{equation}
where $\vec{e}_{v_\infty}$ is the unit vector parallel to the 
initial velocity at infinity,
and $\vec{e}_{b_\infty}$ is the vector perpendicular to $\vec{e}_{v_\infty}$ 
in the plane perpendicular to the one defined 
by $\vec{r_0}$ and $\vec{v_\infty}$, i.e.
the direction of the impact parameter, pointing towards an
increase in $b$.
\begin{equation}
\vec{v_0}=v_0\cos (\alpha -\alpha _0)\vec{e}_{v_\infty}-v_0
\sin (\alpha -\alpha _0)\vec{e}_{b_\infty}
.\end{equation}
The parameters $v_0$, $r_{imp}$ and $\alpha $, $\alpha _0$ are derived
from the analytical expressions of a hyperbolic orbit (see 
L\'opez-Corredoira et al. 2002a; subsect. 3.3) and taking into account
that the angular momentum is constant along the orbit:

\begin{equation}
v_0=\sqrt{\frac{2GM(r_0)}{r_0}+v_\infty ^2}
,\end{equation}
\begin{equation}
r_{imp}=r_0|\sin [\beta -\cos ^{-1}(-1/\epsilon )]|
,\end{equation}
\begin{equation}
\beta =\cos ^{-1}\left[\frac{\left(\frac{\epsilon}{A\ r_0}-1\right)}
{\epsilon }\right]
,\end{equation}
\begin{equation}
A=\frac{v_\infty^2}{GM(r_0)}\frac{\epsilon }{\epsilon ^2-1}
,\end{equation}
\begin{equation}
\epsilon=\sqrt{1+\left(\frac{bv_\infty^2}{GM(r_0)}\right)^2}
,\end{equation}
\begin{equation}
\alpha _0=\frac{\pi }{2}+\cos ^{-1}\left(\frac{r_{imp}}{r_0}\right)
\ \ {\rm (defined\ to\ be\ between\ \pi /2 \ and\ \pi )} 
,\end{equation}
\begin{equation}
\alpha =\pi -\sin ^{-1}\left(\frac{bv_\infty}{r_0v_0}\right)
\ \ {\rm (defined\ to\ be\ between\ \pi /2 \ and\ \pi )} 
.\end{equation}

The results of the orbit with these analytical expressions were compared
with the results of numerical calculations, and they agree. By applying these
formulae, we do not need to do numerical calculations beyond 400--500 kpc.

\subsection{Histogram of energy variations}
\label{.histogram}

The only free parameters for the calculation of the energy variation
of each particle once the galactic parameters
are established are $\rho _\infty$,
$v_\infty $ and $\theta _\infty $. We could also consider
$M_{\rm halo}$ as a third free parameter instead of a fixed
value of $M_{\rm halo}=1.2\times 10^{12}$ M$_\odot $, since the total mass 
of the halo might be variable with the time, precisely because of 
the accreting mechanism proposed in this paper: the loss of energy
of some particles which are trapped when this loss of energy is higher
than
\begin{equation}
E_\infty =\frac{1}{2}mv_\infty ^2
.\end{equation}

\begin{figure}
\begin{center}
\vspace{1cm}
\mbox{\epsfig{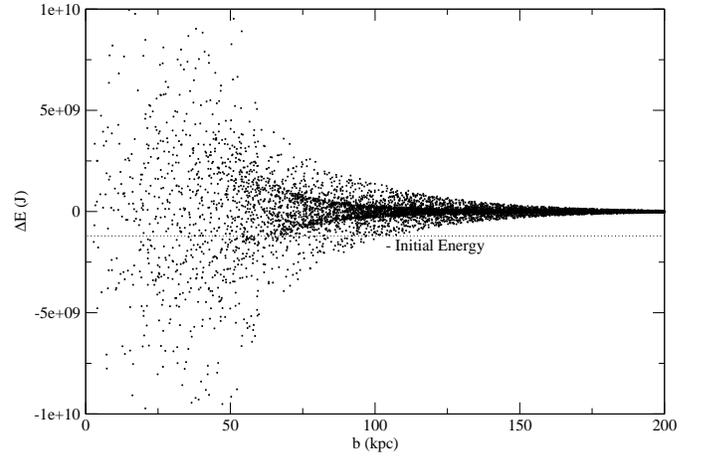}}
\end{center}
\caption{Energy gain/loss in the simulation of particles with
different impact parameter ($b$).}
\label{Fig:histob1}
\end{figure}

\begin{figure}
\begin{center}
\vspace{1cm}
\mbox{\epsfig{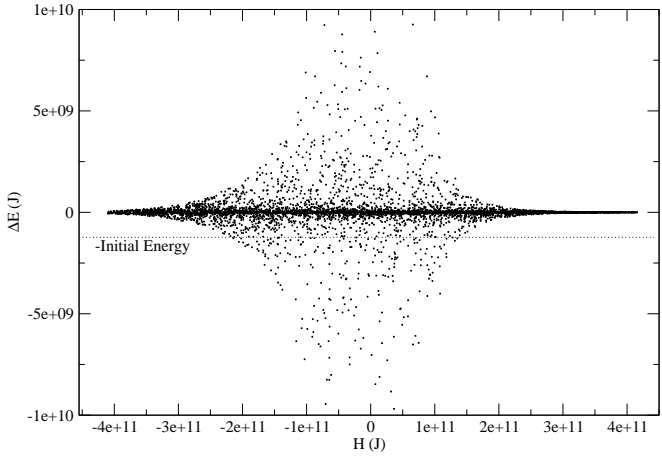}}
\end{center}
\caption{Energy gain/loss in the simulation of particles as
a function of its Jacobi integral. For comparison, the Jacobi integrals 
for the Lagrangian points are 
$H=-1.41\times 10^{11}$ J and $-1.29\times 10^{10}$ J.}
\label{Fig:Hdens}
\end{figure}

As an example, let us set $v_\infty =50$ km/s (a typical value for the peculiar
motion of a galaxy) and $\theta _\infty=45^\circ $, with 
$M_{\rm halo}=1.2\times 10^{12}$ M$_\odot $. The results for 10\,000
particles are in Figs. \ref{Fig:histo1}, \ref{Fig:histob1} and \ref{Fig:Hdens},
which show that the loss of energy is high enough to be trapped
only for $b<100$ kpc or $-2.2\times 10^{11}<H<1.2\times 10^{11}$ J 
[for comparison, the Jacobi integrals 
of the Lagrangian points are 
$H=-1.41\times 10^{11}$ J and $-1.29\times 10^{11}$ J], and
only 426 of the 10\,000 particles lose an energy higher than $E_\infty $. 
Equations (\ref{deltaE}) and (\ref{Fphi}) were
applied to calculated $\Delta E$ for each particle once it moves away
from the galaxy. There are, as observed in Fig. \ref{Fig:histo1},
practically the same number of particles which gain $E\ dE$ as those
which lose $E\ dE$.

Let $R_{200}$ be the ratio of particles within $b<200$ kpc which are
trapped by the halo. According to the above, this is 
$R_{200}=\frac{426}{10000}=0.0426\pm 0.0021$.
Hence, the accretion ratio will be:

\begin{figure}
\begin{center}
\vspace{1cm}
\mbox{\epsfig{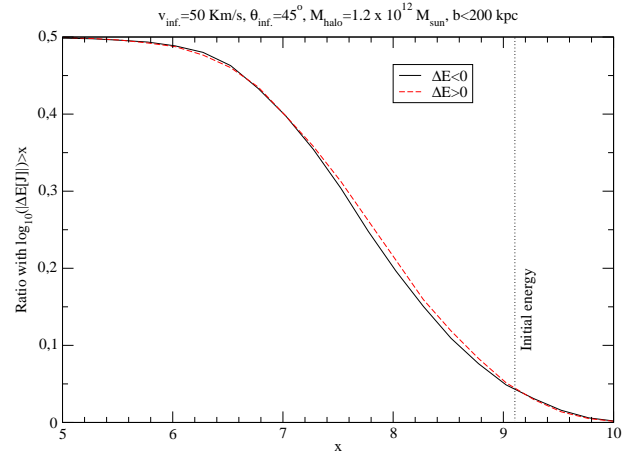}}
\end{center}
\caption{Histogram of distribution of energy gain/loss per unit mass (kg).}
\label{Fig:histo1}
\end{figure}

\begin{equation}
\label{form200}
\frac{dM_{\rm halo}}{dt}=\rho _\infty v_\infty R_{200}\int _0^{2\pi }
d\phi _b\int _0^{200\ {\rm kpc}} 
db\ b\end{equation}\[
=1.9\times 10^{24}\rho _\infty ({\rm kg/m^3})
v_\infty ({\rm km/s}) R_{200}{\rm \ \ M_\odot /yr}
\]\[
=4.0(\pm 0.2)\times 10^{24}\rho _\infty ({\rm kg/m^3}) {\rm \ M_\odot /yr}
\]

\subsubsection{Dependence on the initial infall angle}

\begin{figure}
\begin{center}
\vspace{1cm}
\mbox{\epsfig{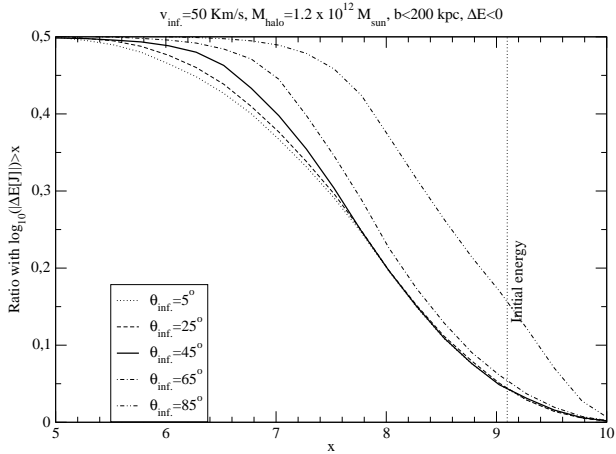}}
\end{center}
\caption{Histogram of distribution of energy loss as a function of
the initial infall angle.}
\label{Fig:histo2}
\end{figure}

The dependence on the angle $\theta _\infty $ is illustrated
in Fig. \ref{Fig:histo2}. The  $R_{200}$ ratios are: 
$0.0441(\pm 0.0021)$ for $\theta _\infty =5^\circ $,
$0.0436(\pm 0.0021)$ for $\theta _\infty =25^\circ $,
$0.0536(\pm 0.0023)$ for $\theta _\infty =65^\circ $,
and $0.1563(\pm 0.0040)$ for $\theta _\infty =85^\circ $.
$R_{200}$ does not change appreciably
except for $\theta _\infty =85^\circ $, which is 3--4 times higher.
The polar accretion seems to favour the mechanism.
The dependence is roughly described by a law:

\begin{equation}
R_{200}(\theta _\infty ; v_\infty=50\ {\rm Km/s}, M_{\rm halo}=
1.2\times 10^{12}\ {\rm M}_\odot )
\end{equation}\[
\approx 0.043+0.20\exp{[-0.12[90- 
\theta _\infty ({\rm deg.})]]}
\]

Since we are interested in calculating the order of
magnitude, we can disregard this dependence and leave for our
remaining calculations a value of $\theta _\infty =45^\circ $.

\subsubsection{Dependence on the initial velocity}
\label{.veldep}

\begin{figure}
\begin{center}
\vspace{1cm}
\mbox{\epsfig{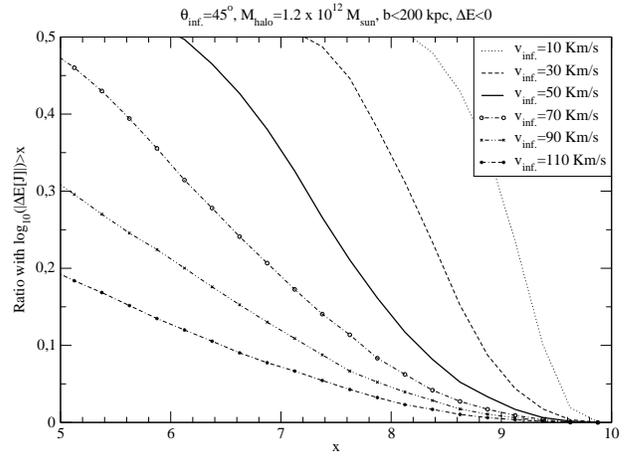}}
\end{center}
\caption{Histogram of distribution of energy loss as a function of 
the initial velocity.}
\label{Fig:histo3}
\end{figure}

However, the dependence on initial velocity is high. 
The histograms in Fig. \ref{Fig:histo3} show it.  
The ratios $R_{200}$ are (note that $E_\infty$ varies with
$v_\infty $): 
$>0.4966$ for $v_\infty =10$ Km/s,
$0.2511(\pm 0.0050)$ for $v_\infty =30$ Km/s,
$0.0118(\pm 0.0011)$ for $v_\infty =70$ Km/s,
$0.0035(\pm 0.0006)$ for $v_\infty =90$ Km/s,
and $0.0012(\pm 0.0003)$ for $v_\infty =110$ km/s.
The ratio of higher energy loss
is maximum for lower initial velocities, both in terms of absolute energy and
relative energy with respect to $E_\infty$. The explanation for
this is multiple. First, for higher velocities the particles go very fast
when they reach the bar, and since $\Delta E$ in equation (\ref{deltaE})
is proportional to the time the particle spends near the bar, the
loss of energy is lower. Second, the capacity to attract  towards the
centre particles with high $b$ is larger for lower values of the initial
velocity. Moreover, $E_\infty$ is higher for higher
initial velocities, so the rate of  energy loss is lower too.
For the cases $v_\infty \le 30$ km/s, it is observed that some particles
with $b>200$ kpc also lose  $\Delta E<-E_\infty$; therefore, the
trapped mass is even higher than the ratio given in eq. (\ref{form200}).
The dependence is roughly  described by the law:

\begin{equation}
R_{200}(v_\infty >\approx 30\ {\rm km/s}; 
\theta _\infty=45^\circ, M_{\rm halo}=
1.2\times 10^{12}\ {\rm M}_\odot )
\end{equation}\[
\approx 0.82\exp{[-0.06v_\infty ({\rm km/s})]}
.\]

\subsubsection{Dependence on the halo mass}

\begin{figure}
\begin{center}
\vspace{1cm}
\mbox{\epsfig{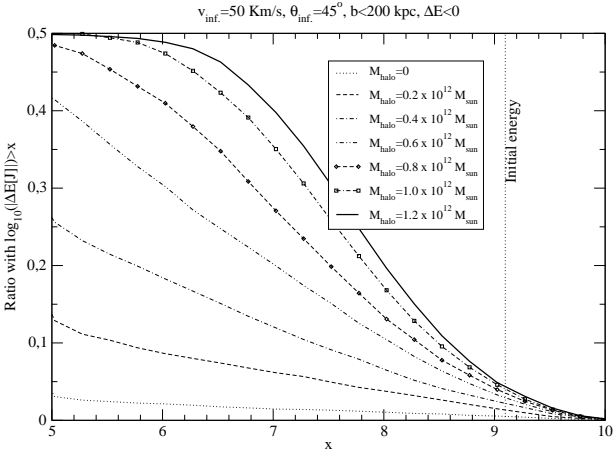}}
\end{center}
\caption{Histogram of distribution of energy loss as a function of 
 halo mass.}
\label{Fig:histo4}
\end{figure}

The effect of the variation of halo mass is similar to the effect
of the variation in the initial velocity because the amount of mass
is proportional to the acceleration, and consequently is related to
the velocity of the particle. Fig. \ref{Fig:histo4} illustrates this.
A maximum accretion is obtained for the largest mass. 
Lower masses have lower capacities to trap particles with high $b$.
The  values of $R_{200}$ obtained are:
0.0049($\pm $0.0007) for $M_{\rm halo}=0$,
0.0092($\pm $0.0010) for $M_{\rm halo}=0.2\times 10^{12}\ {\rm M}_\odot $,
0.0220($\pm $0.0015) for $M_{\rm halo}=0.4\times 10^{12}\ {\rm M}_\odot $,
0.0299($\pm $0.0017) for $M_{\rm halo}=0.6\times 10^{12}\ {\rm M}_\odot $,
0.0362($\pm $0.0019) for $M_{\rm halo}=0.8\times 10^{12}\ {\rm M}_\odot $,
and 0.0396($\pm $0.0020) for $M_{\rm halo}=1.0\times 10^{12}\ {\rm M}_\odot $.
It is clear that it is proportional to the halo mass, roughly according
to:

\begin{equation}
R_{200}(M_{\rm halo}; \theta _\infty =45^\circ, v_\infty=50\ {\rm Km/s})
\end{equation}\[
\approx 0.0062+0.034M_{\rm halo}(10^{12}\ {\rm M}_\odot )
\]

\subsection{Other ways of accretion and the escape of halo particles}

The gain or loss of mass in the 
halo due to other mechanisms will not be considered in this paper, but only the mechanism related to
the dynamical friction with the bar. There are other mechanisms
of accretion, such as  friction with the disc of particles or
clouds that cross it. There are also ways for the halo to lose mass,
such as galaxy stripping in  interactions with other galaxies or the 
IGM, or the escape of particles when they reach
a velocity larger than  escape velocity due to multiple interactions.
We neglect these other mechanisms because we want to consider
only the effect of the bar in increasing the halo mass independently
of other effects. The reality is more complex, of course, but
here we just consider a toy model for rough estimates of the net
effect of an increase in the accretion of mass in barred galaxies with
respect to non-barred galaxies.

The non-conservative potential of the bar could also produce an
increment of the energy of the halo particles and these could reach
the escape velocity producing some kind of ``evaporation'' of the
halo. The mechanism is the same as the one explained in previous subsections
but the initial conditions will be different from those explained
in \S \ref{.initial}. Indeed, the fact that we get some net accretion
of some IGM particles arises because some of them go through the centre of the
galaxy, but the virialized halo has  totally different initial conditions.

Assuming that the halo particles have an initial velocity corresponding
to a circular velocity plus isotropic dispersion velocities with
a component in the radial direction [given the isotropy, the dispersion
in other directions will be the same; the isotropic representation, 
while more or less correct in the inner
halo, is not correct in the outer halo (Dehnen
\& McLaughlin 2005), but we  consider it here as a rough approximation]

\begin{equation}
\sigma _r(r)\propto [\rho (r) r^\alpha]^{1/3}  
,\end{equation}
$\alpha =35/18$ (Dehnen \& McLaughlin 2005), in agreement with
simulations and analytical solutions of self-gravitating dynamical
equilibriums; with the halo density of eq. (\ref{halodens}), and
$\sigma _r (r=a)=v_c(a)/\sqrt{3}$ (Dehnen \& McLaughlin 2005), where
$v_c(r)$ is the circular velocity, we can perform
again the numerical calculation of the orbits and the increment of
energy according to eqs. (\ref{deltaE}) and (\ref{Fphi}).
The result with 10\,000 halo particles whose initial conditions are
distributed randomly according to the probability densities
of positions and velocities including the dispersions, during
14 Gyr orbit calculation, 
gives the result that none of these particles gets
an increase in energy $\Delta E>\frac{GM(r_0)}{2r_0}$, where $r_0$ is
the initial distance from the centre of the galaxy. This means
that the increase in the halo evaporation due to the bar--halo 
interaction is negligible: $\frac{\Delta M}{M}<\sim 10^{-4}$ during
the whole life of the galaxy.

\subsection{Evolution of the halo mass}

\begin{figure}
\begin{center}
\vspace{1cm}
\mbox{\epsfig{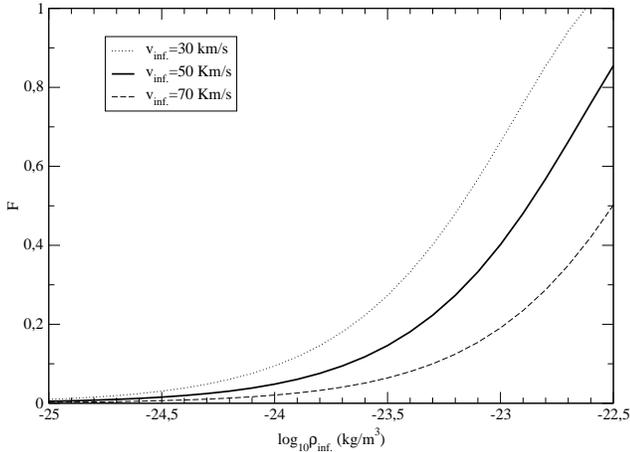}}
\end{center}
\caption{Fraction of halo mass accreted by means of the mechanism 
proposed in this paper.}
\label{Fig:Fdens}
\end{figure}

Let us suppose that the dependence on the initial angle, initial velocity
and the halo mass are separable. This is not strictly correct, but may be appropriate as
a rough estimation . In this case, a generalization
of the expression (\ref{form200}) will be:

\begin{equation}
\frac{dM_{\rm halo}}{dt}=2.1\times 10^{23}\rho _\infty ({\rm kg/m^3})
v_\infty ({\rm km/s})\times 
\end{equation}\[
\exp{[-0.06v_\infty ({\rm km/s})]} 
[1+4.7\exp{[-0.12[90- \theta _\infty ({\rm deg.})]]}]
\]\[
\times    
[1+5.5\times 10^{-12}M_{\rm halo}({\rm M}_\odot )] 
{\rm \ M_\odot /yr}
.\]
Here we are assuming that the disc and the bar do not change with time.
The solution of this differential equation will be:

\begin{equation}
M_{\rm halo}(t)=-1.8\times 10^{11} {\rm M}_\odot +
[1.8\times 10^{11} {\rm M}_\odot +M_{\rm halo}(t=0)]\times e^E
,\end{equation}\[
E=1.2\times 10^{12}\rho _\infty ({\rm kg/m^3})
v_\infty ({\rm km/s})\times \exp{[-0.06v_\infty ({\rm km/s})]}
\]\[
\times 
[1+5.5\exp{[-0.12[90- \theta _\infty ({\rm deg.})]]}]\times t(yr)  
,\]
and the total accretion in the lifetime of the Galaxy 
(we take $t=1.4\times 10^{10}$ yr) due
to this mechanism is
\begin{equation}
\label{deltam}
\Delta M_{\rm halo}=[1.8\times 10^{11} {\rm M}_\odot +M_{\rm halo}(t=0)]
\times [e^E-1]
,\end{equation}\[
E=1.7\times 10^{22}\rho _\infty ({\rm kg/m^3})
v_\infty ({\rm km/s})\times \exp{[-0.06v_\infty ({\rm km/s})]}
\]\[
\times [1+5.5
\exp{[-0.12[90- \theta _\infty ({\rm deg.})]]}]
.\]
The necessary conditions for a fraction $F$ of the halo mass to be accreted
by this mechanism [$M_{\rm halo}(t=0)=(1-F)M_{\rm halo}(t)$, 
$\Delta M=FM_{\rm halo}(t)$; assuming there are no other mechanisms
of accretion] are:

\begin{equation}
F=\left[1+\frac{1.8\times 10^{11} {\rm M}_\odot}{M_{\rm halo}(t)}\right]
\times \left(1+\frac{1}{e^E-1}\right)^{-1}
\label{F}
\end{equation}
Particularly, for $M_{\rm halo}(t)=1.2\times 10^{12}$ M$_\odot $, 
$\theta _\infty =45^\circ $ and $v_\infty=50$ km/s, 

\begin{equation}
F=1.15\left(1+\frac{1}
{\exp{[4.3\times 10^{22}\rho _\infty ({\rm kg/m^3})]}-1}
\right)^{-1}
.\end{equation}
This expression has sense only for $\rho _\infty ({\rm kg/m^3})\le 4.7\times
10^{-23}$ kg/m$^3$ ($F\le 1$), for
higher densities the mass of the halo should be larger.
In Fig. \ref{Fig:Fdens}, the values of $F$ for different densities 
are plotted. Densities of the IGM
around $10^{-24}$ kg/m$^3$ (including baryonic and
non-baryonic matter)
are able to produce a significant amount of 
accretion by means of this mechanism.

In order to have at least 10\% of the total halo matter accreted
by this mechanism ($F=0.1$) it is necessary that (we keep $\theta _\infty =45^\circ $,
$M_{\rm halo}(t)=1.2\times 10^{12}$ M$_\odot $)

\begin{equation}
F>0.1 \Leftrightarrow v_\infty ({\rm km/s})e^{-0.06v_\infty ({\rm km/s})}
\rho _\infty ({\rm kg/m}^3)>5.3\times 10^{-24}
\label{F1}
\end{equation}

\subsection{Considerations about the range of possible values
of the parameters}

\subsubsection{Density of the IGM}

The expression (\ref{F1}) for $F>0.1$ indicates 
that for $\theta _\infty =45^\circ $,
$M_{\rm halo}(t)=1.2\times 10^{12}$ M$_\odot $, $v_\infty =50$ km/s
we need a density of $2\times 10^{-24}$ kg/m$^3=3\times 10^{13}$
M$_\odot $/Mpc$^3=10^2h_0^{-2}\rho _{\rm critical}$. 
This is a high density and  may be reached
by the intracluster gas in some groups or clusters of galaxies,
especially in the richest ones, which can have a density much higher
than this (including dark matter). The typical IGM
densities in clusters are between $10^{-25}$ and $10^{-23}$ kg/m$^3$ (Roussel
et al. 2000). However, the relative velocities 
in such environments are also higher, thus decreasing the desired effect
(see the next subsection). 
For cases such as the Local Group this density is too high.
The average IGM density of the Local Group is $<\sim 10^{-25}$ kg/m$^3$ 
(assuming a total IGM mass of $<\sim 2\times 10^{12}$ M$_\odot $ 
[L\'opez-Corredoira et al. 1999] in a volume of $\sim 1$ Mpc$^3$).

\subsubsection{Velocity of the IGM}

The product $v_\infty e^{-0.06v_\infty}$ is maximum for $v_\infty =17$
km/s, at which a density $9\times 10^{-25}$ kg/m$^3$ produces according
to eq. (\ref{F1}) a $F=0.1$. The density might be even lower because,
as discussed in \S \ref{.veldep}, the accretion of particles with
an impact parameter higher than 200 kpc will be considerable (and we
are considering in our calculation only particles within $b<200$ kpc).
Lower densities are more in the range
of possible values of intracluster density in small groups.
However, the velocity of 10--20 km/s is somewhat low.
For comparison, the velocity of the centre of the Galaxy with respect 
to the barycentre of the Local Group is 81 km/s 
[Braun \& Burton 1999; eq. (3)]. Maybe in the first stages of the
halo formation, when the galactic masses were much lower than
$M_{\rm halo}(t)=1.2\times 10^{12}$ M$_\odot $, the velocities
of the galaxies with respect to the average IGM
were also lower, since the gravitational potential responsible for
such  dynamics produced lower accelerations.
A more developed study is necessary to clarify the question.

In any case, statistically, it is possible to find a significant
fraction of galaxies with low velocities with respect to the average
IGM. The present 
mechanism would apply to these few galaxies.

\subsubsection{Mass of the halo}

Up to now, we have considered the standard model of galaxies
with massive dark matter halos. However, some authors have
questioned the very existence of such halos since its evidence
is weak on galactic scales (Battaner \& Florido 2000; 
Sellwood \& Kosowsky 2001; Evans 2001; etc.) and the rotation
curves or satellite motions can be explained by  alternative 
hypotheses such as magnetic fields (Battaner \& Florido 2000)
or MOND (Sanders \& McGaugh 2002).
If we consider a $M_{\rm halo}(t)=10^{11}$ M$_\odot $,
necessary to explain the rotation curve of the Galaxy up to 15 Kpc
(Honma \& Sofue 1996)
together with a disc + bar mass of $6\times 10^{10}$  M$_\odot $,
we would find from eq. (\ref{F}) that (we keep $\theta _\infty=45^\circ $) 

\begin{equation} 
F>0.1 \Leftrightarrow v_\infty ({\rm km/s})e^{-0.1v_\infty ({\rm km/s})}
\rho _\infty ({\rm kg/m}^3)>8.5\times 10^{-25}
,\end{equation}
in which the necessary density is 6 times lower than in expression
(\ref{F1}).

If the halo mass were still lower than $\sim 10^{11}$ M$_\odot $,
we could reach the conclusion that most of the halo mass was 
accreted by this mechanism ($F$ close to unity). Here, however,
we have not included any alternative effects such as magnetic fields or
MOND, which would be necessary to make compatible the rotation curves
with this low value of the halo.

If the mass of the halo were much higher (Conroy et al. 2005, give
typical masses for the halo of the galaxies of $\sim 5\times  10^{12}$
M$_ \odot$), the relative fraction $F$ would be around 15\% lower than
in expression (\ref{F1}), although,
of course, the amount of accreted matter is nearly
proportional to the halo mass, so in absolute terms the number
of accreted solar masses would be higher. 

\subsection{Strength of the bar}
\label{.barstrength}

In the present paper, we have fixed the parameters of the bar as
given in \S \ref {.parameters} which are typical for a normal strong
bar: $Q_m=5.4\times 10^4$ (in units $10^6$M$_\odot $
kpc$^2$, as in Athanassoula 1992). A weak bar would have
$Q_m\sim 10^4$ (Athanassoula 1992). Since the energy loss is
proportional to $\lambda $ and, $\lambda $ is proportional 
to the mass of the bar [eqs (\ref{deltaE}), (\ref{Fphi})], 
it is straightforward
to deduce that within the linear regime of accretion ($F<<1$)
$F$ will be proportional to mass of the bar. Hence, a weak bar,
typically with a mass five times lower than that used here,
would reduce $F$ by a factor of five.

\section{Other kinds of dynamical friction}
\label{.other}

\subsection{Spiral arms and warps}

As said in \S \ref{.varene}, when the gravitational force varies with time,
we get a non-conservative potential; that is, a potential in which
the energy of the particle in this potential is not constant.
Apart from the bar/triaxial bulge, there are other non-axisymmetric 
structures in the galaxy, which, due to galactic rotation, produce 
this variation in the potential; for instance, the spiral arms or
the warps in the disc. However, these structures   normally have  a mass much
lower than that of the bar (the spiral arm of the Milky Way
would have a $\sim 10$\% of the disc mass
according to the model of Bissantz \& Gerhard 2002; and the mass
of the external disc, where the warp is present [L\'opez-Corredoira
et al. 2002b] is also lower than 10\% of the disc mass); 
 therefore, their effect should be considerably lower, except
perhaps in some cases with prominent spiral arms. These cases
will not be explored in the present paper.

\subsection{Interaction with individual stars}

A way to produce changes in energy is in the interaction of the
accreted particles with individual stars. Since individual stars move, 
the potential of the interaction is not
steady, although it is only predominant with respect to the global
potential of the whole galaxy when the particle approaches too close
to the star. The effect is similar to the exchange of energy of
a comet/spacecraft in an approach to a planet  the Solar System.

\begin{figure}
\begin{center}
\vspace{1cm}
\mbox{\epsfig{file=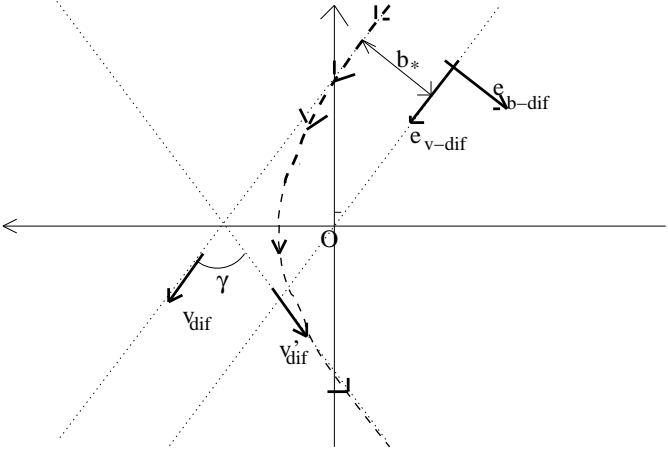,height=6cm}}
\end{center}
\caption{Graphical representation of the hyperbolic trajectory
of a particles which a approach  a star (at the centre O).}
\label{Fig:orbit}
\end{figure}

The velocity of a particle when it approaches  a star of mass 
$M_*$ with respect to this star is 

\begin{equation}
\vec{v_{dif}}=\vec{v}-\vec{v_*}
,\end{equation}
where $\vec{v}$ is the velocity of the particle with respect to the
centre of the galaxy, and $\vec{v_*}$ is the rotation(+peculiar) velocity
of the star around the centre of the galaxy. 
The final velocity of the particle when it 
recedes from the star after the interaction is
\begin{equation}
\vec{v_{dif}}'=\vec{v}'-\vec{v_*}
,\end{equation}
where $\vec{v}'$ is the final velocity with respect to the centre of the
galaxy. The star may be considered point-like and the path
of the particle in the interaction with the star as a hyperbola
with impact parameter $b_*$,
which follows (see subsection 3.3 of L\'opez-Corredoira et al. 2002a)

\begin{equation}
\tan \frac{\gamma }{2}=\frac{G\ M_*}{b_*v_{dif}^2}
\label{gamma}
,\end{equation}
where $\gamma $ is the angle of deviation of the trajectory due
to the interaction as represented in Fig. \ref{Fig:orbit}.

The final velocity with respect to the star will be:

\begin{equation}
|\vec{v_{dif}'}|=|\vec{v_{dif}}|
,\end{equation}
\begin{equation}
\vec{v_{dif}'}=v_{dif}\cos (\gamma )\vec{e_{v-dif}}+
v_{dif}\sin (\gamma )\vec{e_{b-dif}}
,\end{equation}
with the unit vector as specified in Fig. \ref{Fig:orbit},
and with respect to the centre of the galaxy:

\begin{equation}
\vec{v}'=\vec{v_{dif}'}+\vec{v_*}
\end{equation}\[
=(v_{dif}\cos \gamma+v_{*}\cos \alpha ) \vec{e_{v-dif}}+
(v_{dif}\sin \gamma +v_{*}\sin \alpha ) \vec{e_{b-dif}}
,\]\[
\cos \alpha \equiv \vec{v_*} \vec{v_{dif}}
.\]
Hence,
\begin{equation}
v'=\sqrt{v_{dif}^2+v_{*}^2+2v_{dif}v_*\cos (\alpha - \gamma)}
,\end{equation}
while the initial velocity with respect to the centre of the galaxy
is
\begin{equation}
v=\sqrt{v_{dif}^2+v_{*}^2+2v_{dif}v_*\cos (\alpha )}
.\end{equation}
This implies an increment of the energy of the particle
\begin{equation}
\Delta E=\frac{1}{2}m(v'^2-v^2)=mv_{dif}v_*
[\cos (\alpha - \gamma)-\cos (\alpha)]
.\end{equation}
An approximation for small interactions ($\gamma <<1$),
using (\ref{gamma}),
\begin{equation}
\Delta E=\frac{1}{2}m(v'^2-v^2)=2mv_*\sin \alpha 
\frac{G\ M_*}{b_*v_{dif}}
.\end{equation}

The particles trapped by the galaxy due to this interaction will
be those with $\Delta E<-\frac{1}{2}mv_\infty ^2$ ($v_\infty $ as  the velocity of the particle at infinite distance always
is); that is, those
particles with impact parameter 

\begin{equation}
b_*<\frac{4GM_*\sin |\alpha |v_*}{v_{dif}v_\infty }
,\end{equation}
\[ \alpha <0\]
For $M_*=1$ M$_\odot $, $v_*=200$ km/s (the rotation as the main component), 
$v_\infty =50$ km/s,
for an average case of $\alpha =-45^\circ $ and $v_{dif}\sim 100$ km/s,
the constraint is $b_*<\sim 2$ A.U. With the typical surface density
in the galactic disc of $10-100$ star/pc$^2$ (L\'opez-Corredoira et al. 2002b),
the probability of a particle  having an impact parameter less than 2 AU
is $\sim 3-30\times 10^{-9}$. This is just a rough estimate but
the order of magnitude should be more or less correct. Therefore,
the accretion ratio is much lower than the values of $R_{200}$ obtained
in \S \ref{.histogram}, and a much lower accretion (totally negligible)
is expected to be produced in terms of the  star/particle interaction.

\section{Comparison with observations: the effect of the bar on the
halo mass}
\label{.observations}

It is not straightforward to check this effect on real galaxies.
We would need to know the complete history of each galaxy, the epoch
of formation of its bar and other characteristics which are not
directly available. Nonetheless, we can analyse whether  spiral
galaxies with a prominent bar have  different rotation-curve amplitudes from the same type of galaxies without bar.
Indeed, this analysis cannot be a
definitive proof of the present mechanism because there
are other possible alternative explanations in terms of
the interaction between halo and bar. For instance,
halos might allow bars to become stronger
(Athanassoula \& Misiriotis 2002;
Athanassoula 2002, 2003). In any case, the present section might be
a first step in the search of this effect.

In order to do this, we  use first the data from the Mathewson et
al. (1992) catalogue on rotation curves, which contains data
for 1355 spiral galaxies. We  subdivide the sample into 12 groups:
morphological type from +3 to +8 (equivalently from Sb to Sd) and each of
them with or without bar (either SB/SAB or S/SA; according to the
classification in this survey similar to the RC3 catalogue). This is a rough
classification, because indeed most of the spiral galaxies should
have a bar to some degree (e.g., Knapen et al. 2000, Fathi 2004). 
From Mathewson et al. (1992),
we take the maximum rotation velocity. 
This catalogue includes data from optical H$_\alpha $ rotation curves
and HI profiles. The results 
are given in Table \ref{Tab:mathewson}.
The table, although not conclusive, seems to show larger
rotation velocities for earlier types (for +4, the average
rotation velocity of barred galaxies is 2.9$\sigma $ higher than
the non-barred galaxies) and vice verse for later types
(for +7, the average
rotation velocity of barred galaxies is 3.1$\sigma $ lower than
the non-barred galaxies). The first thing might be explained
in terms of the mechanisms proposed in this paper: throughout their lifetimes, bars
help to accrete matter and this is more
remarkable in earlier types because the bars are more prominent
in these types. The hypothesis that halos allow bars to become
stronger is also a good possible explanation; indeed, the simulations in
Athanassoula \& Misiriotis (2002) and Athanassoula (2002, 2003) 
refer only to early-type disc galaxies, with no or
hardly any gas, and this would explain why this effect is not
necessarily seen in later types.
The second fact is intriguing: it might
have to do with the history of the galaxy, in which these late-type
galaxies without bars had  one in the past, or with a major role of the
spiral arms in non-barred galaxy. Also, errors in the classification
of the galaxy type could be the cause of this different velocity
in barred and non-barred galaxies (the bar could contribute to wrong
visual classification).

HI profiles with extension of the wings up to a 20\%
of the maximum trace the halo mass within larger radius, so
the width of this HI profile ($HW20M_{HI}$ in Table
\ref{Tab:mathewson}) is also considered. 
The results are similar to the results
with the rotation curves for late-type spiral galaxies, and
no trend is found for early-type ones.

\begin{table*}
\caption{Comparison of $v_{\rm max}$ [km/s] (maximum velocity of 
rotation curves) and $HW20M_{HI}$ [km/s]
(Half of the HI profile between points where
the intensity falls to 20\% of the highest
channels in each half of the profile and corrected for relativistic effects;
Note: average only with the galaxies which have this datum available)
for spiral galaxies in Mathewson et al. (1992) catalogue in barred and
non-barred cases as a function of the galaxy type.}
\label{Tab:mathewson}
\begin{center}
\begin{tabular}{lll}
Galaxy type & Non-barred: (N), $\langle v_{\rm max}\rangle$,
$\langle HW20M_{HI}\rangle$ & Barred: (N), $\langle 
v_{\rm max}\rangle $, $\langle HW20M_{HI}\rangle$ \\ \hline

+3 & (339), $174\pm 3$, $164\pm 5$ & (23), $182\pm 15$, $180\pm 17$ \\

+4 & (114), $155\pm 5$, $156\pm 7$ & (10), $205\pm 17$, $158\pm 20$ \\

+5 & (96), $127\pm 6$, $127\pm 9$ & (18), $132\pm 13$, $125\pm 13$ \\

+6 & (526), $150\pm 2$, $150\pm 3$ & (31), $127\pm 8$, $116\pm 6$  \\

+7 & (26), $112\pm 9$, $117\pm 13$ & (8), $77\pm 7$, $81\pm 7$ \\

+8 & (55), $91\pm 4$, $96\pm 4$ & (11), $78\pm 5$, $85\pm 5$ \\

\end{tabular}
\end{center}
\end{table*}

The same comparison between barred and non-barred galaxies could
be useful if made in galaxies in clusters. There are very
few galaxies of Mathewson et al. (1992) classified as SB 
and ``in cluster'': according to the 
SIMBAD database (http://simbad.u-strasbg.fr/), only 5 of the 95
barred galaxies are known to be in clusters (`GiC'), so this
catalogue is not useful for this purpose.
We can use another catalogue with fewer galaxies in total (329 galaxies)
but with a higher ratio of cluster members: Vogt et al. (2004).
Results are shown in Table \ref{Tab:vogt} for all the galaxies,
and in Table \ref{Tab:vogt} only for galaxies in clusters.
Nothing new is observed with respect to the Mathewson et al. (1992)
analysis, which shows more or less the same trends. Analysis of 
barred galaxies in clusters (there are in total 23 for our analysis;
not many, but more than the 5 galaxies in Mathewson et al. 1992)
does not show a clearer trend of distinction with the non-barred galaxies:
table \ref{Tab:vogt2}.

\begin{table*}
\caption{Comparison of $OW3$ [km/s] (measured velocity width in optical
profiles with corrections for the shape of the rotation curve,
cosmological stretch and converted to edge-on viewing measured
at twice the scale length), $RW2$ [km/s] (measured 21 cm line profile velocity 
corrected and converted to edge on viewing;
Note: average only with the galaxies which have this datum available), 
$M/L$ [solar units] (corrected mass to light [I-band] ratio;
Note: average only with the galaxies which have this datum available)
for spiral galaxies in Vogt et al. (2004) catalogue in barred and
non-barred cases as a function of the galaxy type. There are no
barred galaxies with available data in this catalogue of types +6 or
later.}
\label{Tab:vogt}
\begin{center}
\begin{tabular}{lll}
Galaxy type & Non-barred: (N), $\langle OW3\rangle$,
$\langle RW2\rangle$, $\langle M/L\rangle$ 
& Barred: (N), $\langle OW3\rangle$,
$\langle RW2\rangle$, $\langle M/L\rangle$
\\ \hline

+3 & (94), $353\pm 12$, $388\pm 14$, $0.47\pm 0.03$ & (25), $404\pm 26$, 
$434\pm 35$, $0.44\pm 0.06$ \\

+4 & (37), $321\pm 16$, $359\pm 21$, $0.52\pm 0.04$ & (3), $419\pm 36$, 
$391\pm 57$, $0.76\pm 0.09$ \\

+5 & (83), $303\pm 11$, $331\pm 10$, $0.46\pm 0.03$ & (9), $252\pm 28$, 
$295\pm 43$, $0.52\pm 0.11$ \\

\end{tabular}
\end{center}
\end{table*}

\begin{table*}
\caption{Same as Table \protect{\ref{Tab:vogt}} but only with galaxies
considered  bona fide members of a cluster of galaxies. There is
only one barred galaxy of type +4 ``in a cluster'' of Vogt et al. (2004) 
catalogue, which is not included.}
\label{Tab:vogt2}
\begin{center}
\begin{tabular}{lll}
Galaxy type & Non-barred: (N), $\langle OW3\rangle$,
$\langle RW2\rangle$, $\langle M/L\rangle$ 
& Barred: (N), $\langle OW3\rangle$,
$\langle RW2\rangle$, $\langle M/L\rangle$
\\ \hline

+3 & (73), $350\pm 12$, $393\pm 15$, $0.47\pm 0.03$ & (19), $384\pm 31$, 
$383\pm 32$, $0.40\pm 0.07$ \\

+5 & (33), $317\pm 17$, $346\pm 16$, $0.49\pm 0.05$ & (4), $234\pm 55$, 
$321\pm 91$, $0.46\pm 0.19$ \\

\end{tabular}
\end{center}
\end{table*}

It would be better to obtain information on the total mass of
halos rather than the mass within a few tens of kpc, as  is the
case for rotation curves, but this information is not available
for a large sample of galaxies, and the errors in the estimates
(through satellites orbit measures; e.g. Conroy et al. 2005) are 
inaccurate and not available for a large sample of individual galaxies to
do statistics.

\section{Conclusions and discussion}

A new mechanism has been proposed in the interaction of bars with
the surrounding medium, which
produces a variation of energy of the particles  crossing
the galaxy. On average, the  energy gained is equal to the loss
energy by these particles, so this does not produce variations
of energy in the average halo. However, for the external IGM
particles which cross the galaxy, this mechanism causes some
of them to be trapped to form part of the halo because their
velocities are reduced to  less than the escape velocity.
By means of this, an extra amount of IGM matter can
be accreted onto barred galaxies.
The increase in the evaporation of the halo due to bar
dynamical friction, however, is  negligible.

Rough calculations of the amount of accreted matter show that this
must be negligible in most cases with normal expected conditions
of density and galaxy--IGM velocity, and only in galaxies of
galaxy--IGM velocities  $<\approx 70$ km/s 
with IGM density similar to that of the cluster of galaxies does it
produce significant ($>10$\%) amounts of halo mass accreted by
this mechanism. The IGM density might be lower
with motions of the galaxies close to the IGM average motion,
or with motion perpendicular to the plane of the galaxy.
Clusters of galaxies  also have other mechanisms
of galaxy stripping that can counteract the accretion; indeed,
the IGM and halos are mixed and we could even talk about a
common halo for all the galaxies in a cluster. This means that we might find
that there is no important increase  in mass in the barred galaxies of
clusters.

The analysis of observational data of rotation curves
shows no clear trends. 
There is a slight increase in the average mass of early-type barred
spiral galaxies with respect to  non-barred galaxies of the same type,
and viceverse for late-type spiral galaxies, but some sort of selection
factor might explain it. No effect could be observed for galaxies
embedded in clusters, although we cannot exclude the possibility that the effect
proposed in this paper is present to some non-negligible degree 
because it might be lower
than the error bars. In Tables \ref{Tab:mathewson}, \ref{Tab:vogt} 
and \ref{Tab:vogt2} the relative error bars of
the velocities are higher than  10\%, which implies a mass error 
 higher than 20\%, so possible slight average variations of
less than $\sim 20$\% in mass would not be detectable.

The fashion in galactic models nowadays is to include very massive halos whose
matter comes from the evolution of the initial fluctuations of
the large scale structure (e.g., Betancort-Rijo \& L\'opez-Corredoira
2002) and the accretion of dwarf galaxies, the hierarchical scenario 
(e.g., White \& Frenk 1991). The mechanism presented in this paper is part of
the possible mechanisms of accretion and it can also be applied
to the satellite dwarf galaxies. Nevertheless, if the very massive
dark matter halos did not exist and the rotation curves were explained
by mechanisms different  from dark matter (e.g., MOND, Sanders \& McGaugh
2002; magnetic fields, Battaner \& Florido 2000), the present
mechanism could be much more important as a relative contributor to
the accretion of matter onto the galaxy, and it would not be detected
by analysing  rotation curves because these
would not be related to the halo mass.

What is noteworthy in this paper is that the present mechanism does
not suppose any speculative scenario about the galaxies themselves.
The mechanism should be present ``always''. The only question is
whether the amount of accreted matter in barred galaxies
is significant enough to be detected or not, because this depends on the
conditions of the IGM.

\ 

{\bf Acknowledgments:} 
Discussions with J. E. Beckman (IAC, Tenerife, Spain) have produced
some ideas which were used as a basis for the development of the present
paper. Thanks are given to E. Athanassoula (Obs. Marseille, France)
and the anonymous referee
for helpful comments on the draft of this paper and suggestions to
improve it; and T. J. Mahoney (IAC, Tenerife, Spain) for proof-reading
of this paper. This research has made use of the SIMBAD database, 
operated at CDS, Strasbourg, France.

\end{document}